\documentclass[11pt,a4paper,twocolumn]{book}
\usepackage{jnpcs}
\usepackage{graphics} 
\usepackage{epsfig}
\newcommand{\beq}{\begin{equation}}
\newcommand{\eeq}{\end{equation}}
\newcommand{\beqa}{\begin{eqnarray}}
\newcommand{\eeqa}{\end{eqnarray}}
\newcommand{\ba}{\begin{array}}
\newcommand{\ea}{\end{array}} 

\ntitle{Multi Solitons of a Bose-Einstein Condensate \\
in a Three-Dimensional Ring}

\nauthor{L. Salasnich}

\naddress{CNR-INFM, Research Unit of Milano University, \\
Dipartimento di Fisica, Universit\`a di Milano, \\
      \it   via Celoria 16, 20133 Milano, Italy \\
{\small \rm
   E-mail: salasnich@.mi.infm.it } \\ 
}

\ndata{
(Received 30 October 2005) }

\nabstract{ 
A Bose-Einstein Condensate (BEC) made of alkali-metal atoms 
at ultra-low temperatures is well described by
the three-dimensional cubic nonlinear Schr\"odinger 
equation, the so-called 
Gross-Pitaevskii equation (GPE). Here we consider an attractive BEC in a ring 
and by solving the GPE we predict the existence of bright solitons 
with single and multi-peaks, showing that they have a 
limited domain of dynamical stability. We also discuss 
finite-temperature effects on the transition from 
the uniform phase to the localized phase.}

\nkeywords{Matter waves; Solitons in Bose-Einstein condensates}

\nPACSnumbers{03.75.-b; 03.75.Lm}
\begin{document}
\DeclareGraphicsExtensions{.jpg,.pdf,.mps,.png} 
\firstpage{1} \nlpage{2} \nvolume{4} \nnumber{1} \nyear{2006}
\def\nfpage{\thepage}
\thispagestyle{myheadings} \npcstitle

Single and multiple bright soliton configurations in 
a Bose Einstein condensates (BEC) 
have been observed in two different experiments \cite{r1,r2}. 
These solitons, which can travel 
for long distances without dispersion, 
have been theoretically investigated by various authors due to 
their relevance in nonlinear atom optics \cite{r3,r4}. 
Very interesting appears the not yet experimentally studied 
case of an attractive BEC in a ring. For this system 
it has been predicted, on the basis of the 
one-dimensional (1D) theory, a quantum phase transition from 
a uniform condensate to a bright soliton \cite{r5}. 
\par 
Here we consider an attractive BEC in a 3D ring, 
taking into account transverse variations of the BEC width. 
We show that the phase diagram of the system reveals novel 
and peculiar phenomena: the localized soliton has a limited 
existence and stability domain and the system supports also multi-peak 
solitons, which can be dynamically stable. 
\par 
The time-dependent Gross-Pitaevskii equation (GPE) 
for a BEC in a trapping potential $U({\bf r})$ is given by 
$$
\left[ i\hbar \frac{\partial}{\partial t} 
+{\hbar^2 \over 2 m} \nabla^2 
- U({\bf r}) -  {4 \pi \hbar^2 a_s (N-1)\over m} 
|\psi|^2 \right] \psi = 0 \;  
$$
where $N$ is the number of condensed atoms 
and $a_s$ the s-wave scattering length of inter-atomic 
interaction. The GPE describes $N$ trapped bosons in the same 
quantum state $\psi({\bf r},t)$: a pure BEC of $N$ atoms. 
The GPE is a mean-field Hartree equation and it is accurate 
for a dilute gas of condensed bosons at ultra-low temperature. 
The cubic term is obtained by using a contact pseudo-potential 
$V({\bf r},{\bf r}') =  {4 \pi \hbar^2 a_s /m} \; 
\delta^3({\bf r}-{\bf r}')$ 
to model the two-body interaction. If $a_s>0$ there is 
repulsion between particles, if $a_s<0$ there is attraction between 
particles \cite{r6}. 
\par 
We describe an attractive BEC in a cylinder, i.e. 
under transverse radial harmonic confinement and 
without axial longitudinal confinement, by using 
the self-focusing GPE ($a_s<0$) with $\omega_{\bot}$ 
the transverse frequency of harmonic confinement. 
This self-focusing 3D GPE is the Euler-Lagrange equation 
of the following Lagrangian density 
$$
{\cal L} = \psi^*   
\left[ i {\partial \over \partial t} + {1\over 2} \nabla^2 
- \frac{1}{2} (x^2 + x^2) + {1\over 2}\Gamma |\psi|^2 \right] \psi \;  
$$
where $\Gamma = 4 \pi (N-1) |a_s|/a_{\bot}$ is the scaled 
inter-atomic strength. Here we use scaled units: energy in units 
$\hbar\omega_{\bot}$, time in units of $\omega_{\bot}^{-1}$, 
length in units of $a_{\bot}=\sqrt{ \hbar / (m\omega_{\bot}) }$. 
To simplify the 3D problem, let us choose the trial wave function 
$
\psi({\bf r},t) = f(z,t) 
{ \exp{\left[ -(x^2+y^2)/2 \sigma(z,t)^2 \right]} 
/ \pi \sigma(z,t)^2  } 
$. 
Inserting it into the Lagrangian, integrating over $x$ and $y$ 
and neglecting the derivatives of $\sigma(z,t)$ one finds 
an effective Lagrangian density ${\bar{\cal L}}$
which depends on the axial wave function $f(z,t)$ 
and the transverse width $\sigma(z,t)$. 
The two Euler-Lagrange equations of the effective Lagrangian 
${\bar{\cal L}}$ with respect to $f$ and $\sigma$ are 
$$
i {\partial \over \partial t} f = 
\left[ -{1\over 2} {\partial^2\over \partial z^2}  
+ {1\over 2} ({1\over \sigma^2} + \sigma^2 ) 
- {g \over \sigma^2} |f|^2 \right] f \;  
$$
$$
\sigma^2 = \sqrt{1 - g |f|^2 } \;   
$$
where $g = \Gamma /(2 \pi) = 2 (N-1) |a_s|/a_{\bot}$. 
For $\sigma=1$ the first equation becomes the familiar self-focusing 
1D GPE, also called cubic nonlinear Schr\"odinger equation.  
The equation for $\sigma$ can be inserted into the equation for $f$. 
In this way one gets the time-dependent non-polynomial Schr\"odinger 
equation (NPSE) \cite{r7}. 
\par 
In \cite{r7} we have verified that the NPSE gives results always 
very close to that of the 3D GPE under transverse harmonic confinement 
for both positive and negative scattering length $a_s$. 
The self-focusing NPSE ($a_s<0)$ admits a shape-invariant solution,  
called bright soliton, due to the interplay between 
the dispersive kinetic term and the attractive interaction term. 
Note that bright solitons of the self-focusing 
3D GPE do not exist without transverse confinement. 
We have found \cite{r3,r4} that 
the bright soliton reduces, in the limit of very 
strong transverse confinement, namely for $g<<1$ (1D regime),  
to the text-book bright soliton of the self-focusing 
1D GPE, whose axial wave function is 
$
f(z) = \sqrt{g/ 4} \; sech({g z/ 2}) \;   
$
Contrary to the self-focusing 1D GPE, 
the bright soliton of the self-focusing NPSE 
has a limited domain of existence: 
$
0 < g < {4\over 3}  
$. 
If the strength $g$ exceeds $4/3$ there is the so-called 
collapse of the BEC. Thus, the transverse dynamics of the BEC induces  
the instability of the solitonic solution.  
\par 
As previously stressed, 
bright solitons have been observed in two experiments: 
a single soliton in expulsive trap at 
the Ecole Normale Superieure (ENS) of Paris 
\cite{r2}, and a soliton train in a harmonic trap 
at Rice University \cite{r1}. We belive that 
in the next future there will be experiments on attractive 
BEC in a toroidal trap. 
In fact, toroidal traps have been recently produced: 
at Georgia Tech cold atoms in a magnetic ring  
(diameter of $2$ cm) \cite{r8} and at Berkeley a 
BEC in a magnetic ring (diameter of $3$ mm) \cite{r9}. 
\par 
To model an attractive BEC in a toroidal trap we consider now 
the self-focusing NPSE where the solution $f(z,t)$ has periodic 
boundary conditions: 
$$
f(z+L,t)=f(z,t) \;   
$$
where $L=2\pi R$ is the azimuthal longitudinal length of the BEC 
and $R$ is the azimuthal radius of the trap. 
Due to the periodic boundary conditions the stationary self-focusing 
NPSE admits a uniform solution, given by 
$
f(z)={1/ \sqrt{L}}  
$.  
One sees immediately that the uniform solution 
exists only for $g < 2\pi R$. The stationary NPSE admits also 
nodeless and nodal localized solutions (solitonic solutions). 
In Fig. 1 we plot the existence diagram of these solitonic solutions 
in the $(R,g)$ plane. The diagram is obtained by numerically solving the 
NPSE. The one-peak localized nodeless solution 
exists between the two solid lines of Fig. 1. 
Nodeless two-peak localized solution
exists between the two dashed lines while the nodal two-peak
localized solution exists between the two dot-dashed lines. 

\begin{figure}[ht!]
     \leavevmode
\centering
\includegraphics[width=2.8in, height=2.5in, angle=0]{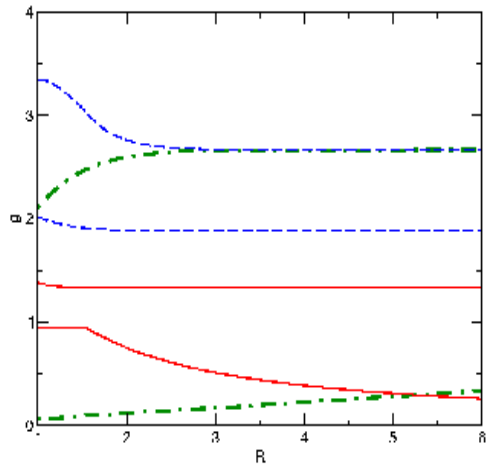}
\caption{Existence diagram in the $(R,g)$ plane, where
$g=2N|a_s|/a_{\bot}$ is the interaction coupling
and $R=L/(2\pi )$ is the azimuthal radius of the ring (in units $a_{\bot}$).
Uniform solution: exists for $g < 2\pi R$.
One-peak localized nodeless solution: exists between the
two solid lines. Nodeless two-peak localized solution
exists between the two dashed lines. The nodal two-peak
localized solution exists between the two dot-dashed lines.}
\end{figure}

We analyze the dynamical stability of the solutions by numerically solving  
the time-dependent NPSE taking as initial condition 
the stationary localized solution $f(z)$ with a very weak 
perturbation. In the left panels of Fig. 2 we plot the 
density profile $\rho(z)=|f(z)|^2$ of the solitonic solutions. 
In the right panels we show the time-dependence 
of the mean squared widths $\langle z^2 \rangle$ 
for the weakly perturbed solitonic solutions. 
A periodic oscillation means that the solitonic solution is 
dynamically stable. In the calculations we choose a 
ring axial lenght $L=15$ and an interaction strength $g=1$. 
We have studied the dynamical stability for various 
initial conditions. The main results are the following: 
i) the uniform solution is stable only below the lower curve 
of existence of the one-peak soliton; 
ii) the one-peak soliton is stable where it exists; 
iii) the nodal two-peak soliton is 
stable in the plane $(R,g)$ only below the upper 
curve of existence of the one-peak soliton; 
iv) the nodeless two-peak soliton is unstable.  
Similar results are found for solitonic 
solutions with a larger number of peaks. 

\begin{figure}[ht!]
     \leavevmode
\centering
\includegraphics[width=3.in, height=2.6in, angle=0]{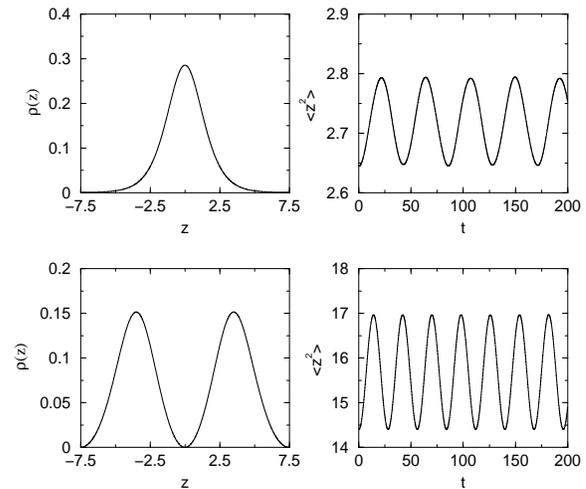}
\caption{
Left panels: density profile $\rho(z)$ 
of the solitonic solutions. Right panels: time-dependence
of the mean squared widths $\langle z^2 \rangle$
for the weakly perturbed solitonic solutions.
Ring axial lenght: $L=15$. Interaction strength: $g=1$.}
\end{figure} 

To conclude, we observe that the critical strength 
$g^*=\pi a_{\bot}/2R$ of the transition from the uniform 
state to the localized state at zero temperature has been determined 
in \cite{r5} by investigating the Bogoliubov energy spectrum of the 
the self-focusing 1D GPE. The inclusion of finite-temperature 
effects is quite simple if one consider a non-interacting thermal 
cloud. In this way we find that the critical temperature $T^*$ 
of transition from the uniform to the localized state is given by 
$$
{k_B T^*\over \hbar \omega_{\bot} } = {a_{\bot}^2 \over 4R^2} 
\left( N - {\pi a_{\bot}^2 \over 4 R|a_s|} \right) 
$$
where $k_B$ is the Boltzmann constant, and using not-scaled quantities. 
Obviously if $N<\pi a_{\bot}^2/(4 R|a_s|)$ the system remains uniform 
at all temperatures.

\end{document}